\documentstyle[12pt]{article}
\begin{document}
\title{Zamolodchikov-Faddeev Algebra in 2-Component Anyons}
\author{Yue-lin Shen and Mo-lin Ge \\
         Theoretical Physics Division, Nankai Institute of Mathematics\\
         Nankai University, Tianjin 300071, China}
\maketitle
\vspace{2cm}
\begin{abstract}
\baselineskip24pt
   We investigate Wilczeck's mutual fractional statistical model at the field-
   theoretical level. The effective Hamiltonian for the particles is derived
    by the canonical procedure, whereas the commutators of the anyonic
    excitations are proved to obey the Zamolodchikov-Faddeev algebra.
    Cases leading to well
    known statistics as well as Laughlin's wave function are discussed.\\
\vspace{3cm}
\end{abstract}
\newpage
\baselineskip24pt
\newcommand{\x}{\mbox{$\bf x$}}
\newcommand{\y}{\mbox{$\bf y$}}
\newcommand{\z}{\mbox{$\bf z$}}
     Fractional statistics plays an important role in planar physics. In two
  dimensional space identical particles can obey new kinds of statistics
   ~\cite{Wi1}, interpolating between the normal Bose and Fermi Statistics.
   The general theory of fractional statistics was put forward by     Wilczeck
and
 Zee ~\cite{WZ}  and Wu ~\cite{Wu1} who interprete the theory in the path
integeral
     formalism and  related the theory to the braid group, while the connection
with the
   Chern-Simons theory was also proposed ~\cite{Wu2,ASWZ}. Since then, a new
    subject, called anyonic physics, has been widely developed, especially in
  the application in High-$T_{c}$ superconductivity  ~\cite{Wi2} and quantum
Hall
 effect ~\cite{PG}.

    Recently Wilczeck has proposed a 2-components anyonic model to describe
 incompressible liquid quantized Hall states in situations where two distinct
 kinds of electrons are relevant. The model can also overcome the violation
 of the discrete symmetries P and T in the original anyonic superconductivity
theorem. In his paper ~\cite{Wi3} mutual statistics between distinguished
particles are introduced, which brings about some interesting results.

    Multianyon quantum mechanics poses a challenging problem in theoretical
     physics. A complete solution is far out of research, we can only
understand
      some features from different point of view.  One of them is to know the
       statistical character of the anyonic excitation. In this work, based on
the
  second quantized procedure, we diagonalize the Hamiltonian for the
2-components
  anyons, at the expense of the unusual commutation relation of the field
   operators, which shares the Zamolodchikov-Faddeev algebra. Though the method
    here is not developed first time, the result has never appeared in the
   literature. The algebra of  these operators   reflects the    statistical
    characters between the anyonic excitations. Special values for the
coupling constans  leading to the Bose, Fermi and q-deformed statistics are
discussed
,meanwhile the Laughlin's wave function is given.

    Let us start from Wilczeck's model ~\cite{Wi3}
    \begin{eqnarray}
    L&=&\sum_{i,a=1,2}\{{1\over 2}M(\frac{d\x_{i}^{(a)} }{dt})^{2}+{e\over c}
  a^{(a)}(\x_{i})\frac{d\x_{i}^{(a)}}{dt}-ea_{0}^{(a)}(\x_{i})\}  \nonumber \\
     & & +{1\over 2\pi} \int\, d\x\, n_{ab}\epsilon^{\alpha\beta\gamma}
     a_{\alpha}^{(a)}f_{\beta\gamma}^{(b)}
    \end{eqnarray}
     where
 \[ f_{\beta\gamma}^{(a)}=\partial_{\beta}a_{\gamma}^{(a)}-\partial_{\gamma}
     a_{\beta}^{(a)} \]
 In Eq.(1) the index i refers to a given particle, $\alpha,\beta,\gamma$
  correspond  to the three-dimensional Lorentz coordinates, $( a)$ represents
the
   two kinds of particles, and all coordinates \x\ or $ \x_{i}$ refers to a two
    dimensional vectors. The matrix
  \[ n_{ab}=\left( \begin{array}{cc}
    m_{1} & n    \\
     n    & m_{2}
 \end{array}  \right) \]
 describes the mutual statistics between the two kinds of particles and their
self-
 statistics.

  Since the time components of the gauge fields are Langrangian multiplies, one
   proceeds by eliminating them by varying the Langrangian with respect to them
( $\delta L/\delta a^{0}=0 $). Then one gets
  \begin{equation}
  e\sum_{i}\delta(\x-\x_{i}^{(a)})=\sum_{b=1,2}{n_{ab}\over \pi}(\partial_{1}
  a_{2}^{(b)}-\partial_{2}a_{1}^{(b)})
  \end{equation}
  Due to the two-dimensional identity $\nabla (\x/\x^{2})=2\pi\delta(\x)$, and
   choosing  the Coulomb gauge $\nabla a(\x)=0$, we can suppose the solution of
(2)  as
   \begin{eqnarray}
   a^{(1)}(\x)&=&\sum_{i}{\theta_{11}\over 2\pi}\frac{k\times(\x-\x_{i}^{(1)})}
   {(\x-\x_{i}^{(1)})^{2}}+\sum_{i}{\theta_{12}\over
2\pi}\frac{k\times(\x-\x_{i}^{(2)})}
   {(\x-\x_{i}^{(2)})^{2}}  \nonumber \\
   a^{(2)}(\x)&=&\sum_{i}{\theta_{21}\over 2\pi}\frac{k\times
(\x-\x_{i}^{(1)})}
   {(\x-\x_{i}^{(1)})^{2}}+\sum_{i}{\theta_{22}\over 2\pi}\frac{k\times
(\x-\x_{i}^{(2)})}
   {(\x-\x_{i}^{(2)})^{2}}
   \end{eqnarray}
   where k is the unit vector perpendicular to the plane.

     The assumption turns out to be true when the parameters $\theta_{ij}$
      satisfy
  \begin{eqnarray}
  e\pi&=&m_{1}\theta_{11} +n\theta_{21}  \nonumber \\
  e\pi&=&m_{2}\theta_{22}+n\theta_{12}  \nonumber \\
     0&=&m_{1}\theta_{12}+n\theta_{22}  \nonumber \\
     0&=&m_{2}\theta_{21}+n\theta_{11}
   \end{eqnarray}
   whose solution reads
   \begin{eqnarray}
   \theta_{11}&=&\frac{m_{2}e\pi}{m_{1}m_{2}-n^{2}} \nonumber \\
   \theta_{12}&=&\frac{-ne\pi}{m_{1}m_{2}-n^{2}}    \nonumber  \\
   \theta_{21}&=&\frac{-ne\pi}{m_{1}m_{2}-n^{2}}  \nonumber \\
   \theta_{22}&=&\frac{m_{1}e\pi}{m_{1}m_{2}-n^{2}}
   \end{eqnarray}

     After the second quantization the effective Hamiltonian for the particles
reads
      \begin{equation}
     H=\sum_{a}\,\int d\x{1\over 2M}[(-i\nabla-{e\over
c}a^{(a)}(\x))\Psi^{(a)}(\x)]^{\dagger}
     [(-i\nabla-{e\over c}a^{(a)}(\x)\Psi^{(a)}(\x)]
     \end{equation}
     where
  \begin{equation}
   a^{(a)}(\x)=\sum_{b=1,2}{\theta_{ab}\over 2\pi}\nabla\int d\y\,\theta(\x-\y)
   \Psi^{(b)\dagger}(\y)\Psi^{(b)}(\y)
   \end{equation}
 and $\theta(\x-\y)$ is the azimutal angle of the vector from \y\ to \x\,
which is usually multivalued function. Since  we are  only interested in the
effect of
 interchanging a pair of particles once  ~\cite{Fradkin} to see the statistics,
the first
  sheet of the complex angle is concerned here, informations from higher ones
can be
   trivially derived ~\cite{KKS}.

 Using the Jordan-Wigner transformation ~\cite{Fradkin,IS,KKS}, we can turn
     the interacted Hamiltonian to a  free one accompanying with the
complicated
     fields operators, or anyonic operators. Study on these operators reveals
the
     statistics of the anyonic excitation mode. The  transformation
     \begin{equation}
     \Psi_{F}^{(a)}(\x)=exp\,(-i\sum_{b}{\theta_{ab}\over 2\pi}\int d\z\,
\theta(\x-\z)
     \Psi^{(b)\dagger}(\z)\Psi^{(b)}(\z))\Psi^{(a)}(\x)
     \end{equation}
  make the Hamiltonian (6)
       \begin{equation}
     H_{F}=\sum_{a}\int\, d\x {1\over 2M}(-i\nabla\Psi_{F}^{(a)}(\x))^{\dagger}
     (-i\nabla\Psi_{F}^{(a)}(\x))
     \end{equation}
     As a note, the free Hamiltonian doesn't mean a non-interacted system,
because
      we will deal with a complicated commutatation algebra.

 By a lengthy but not difficult calculation, we can get the commutators of
  the anyonic operators. For instance, from (8) and using the Baker-
  Cambell-Hausdoff formula, we obtain
   \begin{equation}
   \Psi_{F}^{(1)}(\x)\Psi_{F}^{(1)\dagger}(\y)=\delta_{\x\y}-e^{i{\theta_{11}
  \over 2\pi}(\theta(\y-\x)-\theta(\x-\y))}\Psi_{F}^{(1)\dagger}(\y)\Psi_{F}
  ^{(1)}(\x)
  \end{equation}

  For simplicity, we define the ordering in two dimensional space. Given two
vectors $\x,\y$,
  and their components $\x_{1,2},\y_{1,2}$, we define
  \begin{eqnarray}
  \x>\y&\,\,\,\,\,\,& \mbox{if $\x_{2}<\y_{2}$}    \nonumber \\
       &            &\mbox{or $\x_{2}=\y_{2},\,\, \x_{1}<\y_{1}$} \nonumber \\
  \x=\y &           &\mbox{if $ \x_{1}=\y_{1},\,\,  \x_{2}=\y_{2}$}
  \end{eqnarray}
  Further, we define
  \begin{eqnarray}
  sgn(\x-\y)=\left\{ \begin{array}{ll}
       1 & \x>\y  \\
       0 & \x=\y  \\
       -1 & \x<\y    \end{array} \right\}
 \end{eqnarray}
 With these definitions we get
 \begin{equation}
 \Psi_{F}^{(1)}(\x)\Psi_{F}^{(1)\dagger}(\y)=\delta_{\x\y}-e^{i{\theta_{11}
 \over 2}sgn(\x-\y)}\Psi_{F}^{(1)\dagger}(\y)\Psi_{F}^{(1)}(\x)
 \end{equation}

 The same procedures for other commutators  lead to the following
  relation in a compact form
\begin{eqnarray}
\Psi_{F}^{(i)}(\x)\Psi_{F}^{(j)}(\y)&=&S_{kl}^{ij}(\x-\y)\Psi_{F}^{(k)}(\y)\Psi_{F}^{(l)}(\x) \nonumber \\
\Psi_{F}^{(i)\dagger}(\x)\Psi_{F}^{(j)\dagger}(\y)&=&S_{lk}^{*ji}(\y-\x)\Psi_{F}^{(k)\dagger}(\y)\Psi_{F}^{(l)\dagger}(\x)  \nonumber \\
\Psi_{F}^{(i)}(\x)\Psi_{F}^{(j)\dagger}(\y)&=&S_{lj}^{ki}(\y-\x)\Psi_{F}^{(k)\dagger}(\y)\Psi_{F}^{(l)}(\x)+\delta_{ij}\delta(\x-\y)
\end{eqnarray}
where $S_{kl}^{ij}$ is a $ 4\times 4$ matrix, whose nonzero elements are
    \begin{eqnarray}
   S_{11}^{11}(\x-\y)=-e^{{i\theta_{11} \over 2}sgn(\x-\y)}  \nonumber \\
   S_{12}^{21}(\x-\y)=-e^{{i\theta_{12}\over 2}sgn(\x-\y)}  \nonumber \\
   S_{21}^{12}(\x-\y)=-e^{{i \theta_{21}\over 2}sgn(\x-\y)}  \nonumber  \\
   S_{22}^{22}(\x-\y)=-e^{{i \theta_{22}\over 2}sgn(\x-\y)}
   \end{eqnarray}
It is easy to show that the matrix $S_{kl}^{ij}$ satisfies Yang-Baxter Equation
\begin{equation}
S_{kl}^{ij}(\x-\y)S_{np}^{lm}(\z-\y)S_{qr}^{kn}(\z-\x)=
S_{kl}^{jm}(\z-\x)S_{qn}^{ik}(\z-\y)S_{rp}^{nl}(\x-\y)
\end{equation}

  The algebras (14)  and (16) originally appeared in the $ 1+1$ dimensional
integrable
  field theory ~\cite{ZZ,Faddeev,Kulish}, but it happendly comes into the $2+1$
dimensional
  system, and the spectral parameter in the quantum inverse scattering method
is
  replaced here   by the space-like vector. It is  not clear to interprete the
physical
  consequences of the   algebra, because our system is far from the integrable
system, but from it we can
  find some simple  and solvable cases.  \\
      \newcommand{\ei}{\mbox{$4\pi\times integer$}}
  (1) We have two kinds of free Fermions, if
  \begin{equation}
  \theta_{ij}=\ei
  \end{equation}
  (2)One kind is free Fermion, the other one is q-deformed excitation, if
  \begin{equation}
  \theta_{12}=\theta_{21}=\ei,\,\, \theta_{11}=\ei,\,\,\theta_{22}\neq \ei
  \end{equation}
  (3)Two kinds of q-deformed excitations, and anticommutated each other.If
  \begin{equation}
  \theta_{12}=\theta_{21}=\ei,\,\,\theta_{11}\neq \ei,\,\, \theta_{22}\neq \ei
  \end{equation}
  (4)No free excitation mode if
  \begin{equation}
  \theta_{12}\neq \ei
  \end{equation}
  Correspondingly, if we take $\theta_{ij}=2\pi (2k+1),\,k\in N$,we have the
  Boson or q-deformed excitation mode .

  Finally, introducing two sets of complex coordinates $(z_{i},\bar{z_{i}})$
and $(w_{i},
  \bar{w_{i}})$ for the two kinds of particles, we can discuss the problem in
  the first quantized formalism. In parallel to (8) and (9), we have a
Hamiltonian
  without interaction
  \begin{equation}
  H_{F}^{'}=\sum_{a=1,2}\frac{P_{(a)}^{2}}{2M}
  \end{equation}
  Correspondingly the wavefunction is multivalued. We can construct Laughlin's
wave function ~\cite{PG}
  \begin{equation}
|\Psi\rangle_{F}^{'}=\prod(z_{i}-z_{j})^{\theta_{11}/\pi}(w_{i}-w_{j})^{\theta_{22}/\pi}
  (z_{i}-w_{j})^{\theta_{12}/\pi}f(z_{i},\bar{z_{i}};w_{i},\bar{w_{i}})
  \end{equation}
  where $f(z_{i},\bar{z_{i}};w_{i},\bar{w_{i}})$ is invariant under the
interchanging between any
  two  particles.                      \\

\end{document}